\documentclass[a4paper]{panl}

\usepackage{lmodern,graphics,graphicx,amsmath,amssymb,cite}

\originalTeX

\begin{document}

\issuearea{Physics of Elementary Particles and Atomic Nuclei. Theory}

\title{New solutions of viscous relativistic hydrodynamics}

\maketitle

\author{M.~Csan\'ad$^{\,\, a}$, M.~I.~Nagy$^{\,\, a}$, Ze-Fang~Jiang$^{\,\, b,c,d}$ T.~Cs\"org\H o$^{\,\, e,f}$\\
\from{$^{a}$\,E{\"o}tv{\"o}s Lor{\'a}nd University, H-1117 Budapest, P{\'a}zm{\'a}ny P. s. 1/A, Hungary}\vspace{-3mm}
\from{$^{b}$\,Department of Physics and Electronic-information Engineering, Hubei Engineering University,~Xiaogan 432000,~China}\vspace{-3mm}
\from{$^{c}$\,Key Laboratory of Quark and Lepton Physics, Ministry of Education, Wuhan, 430079, China}\vspace{-3mm}
\from{$^{d}$\,Institute of Particle Physics, Central China Normal University, Wuhan 430079, China}\vspace{-3mm}
\from{$^{e}$\,Wigner RCP, H-1525 Budapest 114, P.O.Box 49, Hungary}\vspace{-3mm}
\from{$^{f}$\,EKU KRC, H-3200, Gy{\"o}ngy{\"o}s, M{\'a}trai \'ut 36, Hungary}
}

\begin{abstract}
Relativistic hydrodynamics represents a powerful tool to investigate the time evolution of the strongly interacting quark gluon plasma created in ultrarelativistic heavy ion collisions. The equations are solved often numerically, and numerous analytic solutions also exist. However, the inclusion of viscous effects in exact, analytic solutions has received less attention. Here we utilize Hubble flow to investigate the role of bulk viscosity, and present different classes of exact, analytic solutions valid also in the presence of dissipative effects.
\end{abstract}

\vspace*{6pt}

\noindent


\vspace*{12pt}

\section{Introduction}
The strongly interacting quark gluon plasma (sQGP), discovered more than a decade ago at RHIC, undergoes various phases throughout its evolution. Starting from an initial stage defined by the energy density deposited by the colliding nucleons, and followed by a quick pre-thermalization stage, the sQGP evolves as a nearly perfect fluid, and then it freezes out to produce hadrons later on (after rescattering and decays) observed by the detectors. The fluid stage can be well described by relativistic hydrodynamics. Besides detailed and realistic simulations (which capture many phases of the mentioned time evolution, such as the initial stage, freeze-out dynamics, rescattering, decays) utilizing a numerical solution of the equations of hydrodynamics, exact and/or analytic solutions are also important. These provide more than a simple ``common sense approach'', and can be utilized to obtain an analytic understanding of the connection between the initial and the final state.

The history of applying relativistic hydrodynamics to describe hadron collisions started with Fermi~\cite{Fermi:1950jd} and Landau~\cite{Landau:1953gs} who also formulated the basic equations we use today. The first historically important solutions are the Landau-Khalatnikov~\cite{Khalatnikov:1954aa} solution and the Hwa-Bjorken solution~\cite{Hwa:1974gn,Bjorken:1982qr}. Since then, many solutions were found, see e.g. the review in Ref.~\cite{deSouza:2015ena}. It turned out that even simple solutions capture many features of the observations made in ultra-relativistic heavy ion collisions~\cite{Csorgo:2003ry,Csanad:2009wc,Csanad:2014dpa}. However, all exact, analytic solutions so far solve the equations of perfect fluid hydrodynamics, which may not be exactly true in the scenario present in heavy ion collision.

%Still lacking: non spherical 3D, accelerating, realistic solutions

%Linearized hydro: perturbations
%Kurgyis, MCs, Universe 3 (2017) no.4, 84
%Shi, Liao and Zhuang, Phys.Rev . C90 (2014) no.6, 064912

%A RECENT CHALLENGE
%Many observing Lévy/Cauchy sources :
%How to reconcile with hydro?
%Boltzmann cutoff
%Rescattering?
%Resonances

\section{Relativistic hydrodynamics}

Relativistic hydrodynamics assumes the local conservation of the energy-momentum tensor $T^{\mu\nu}$. Denoting the fluid four-velocity field with $u^\mu$, the flow is perfect (i.e. there is no viscosity and heat conduction) if the energy-momentum tensor's form is written as:
\begin{align}
T^{\mu\nu}=(\varepsilon+p)u^\mu u^\nu-pg^{\mu\nu},\label{e:em}
\end{align}
where $\varepsilon=u_\mu u_\nu T^{\mu\nu}$ is the energy density in the locally comoving frame, $p$ is pressure and $g^{\mu\nu}$ is the metric tensor, assumed to be of the form $diag(1,-1,-1,-1)$. This equation can be completed by an Equation of State (EoS) $\varepsilon=\kappa p$. In to locally comoving frame (where $u^\mu=(1,0,0,0)$) $T^{\mu\nu}$ is diagonal, having the form $diag(\varepsilon,p,p,p)$. In fact a flow is perfect if it can be transformed to the mentioned diagonal form. In the presence of viscosity and heat conduction, the situation is by far not so unambiguous. The reason for this is that the notion of flow is ambiguous if heat conduction is present, due to the relativistic equivalence of mass and energy and their flows. Using the Eckart frame~\cite{Eckart:1940te} where the fluid velocity indicates the flow of a conserved particle number, the above energy-momentum tensor (in a first order expansion around the perfect fluid case) can be given as
\begin{align}
\label{e:eckart}
T^{\mu\nu} =(\varepsilon+p)u^\mu u^\nu-pg^{\mu\nu} + q^{\mu}u^{\nu} +  q^{\nu}u^{\mu} + \pi^{\mu\nu}.
\end{align}
Here heat flow $q^\mu$ and viscous stress tensor $\pi^{\mu\nu}$ can be given as
\begin{align}
q_\mu        =\;& \lambda(g_{\mu\nu}{-}u_\mu u_\nu)\big(\partial^\nu T - Tu^\rho\partial_\rho u^\nu\big),\\
\pi_{\mu\nu} =\;& \eta\big[(g_{\mu\rho}{-}u_\mu u_\rho)\partial^\rho u_\nu + (g_{\nu\rho}{-}u_\nu u_\rho)\partial^\rho u_\mu\big] + \bigg(\zeta - \frac 2d\eta\bigg)(g_{\mu\nu}{-}u_\mu u_\nu)\partial_\rho u^\rho,
\end{align}
and $\lambda$ represents thermal conductivity, $\eta$ and $\zeta$ represent the shear and bulk viscous coefficients, respectively; furthermore $d$ represents the number of spatial dimensions (i.e. the number of spatial components of all Lorentz vectors). These equations create an environment where the first order viscous effects can be studied effectively even with analytic methods.

Before doing so however, we may ask how the above coefficients ($\lambda$, $\eta$, $\zeta$) depend on the thermodynamic quantities. As we later describe, we only focus on bulk viscosity here, and its dependence on temperature was studied e.g. in Refs.~\cite{Karsch:2007jc,NoronhaHostler:2008ju,Harutyunyan:2018wdk,Ryu:2017qzn}. These provide different dependencies, hence we focus here on the simplest possible cases. One choice could be to assume $\zeta\propto s$ (where $s$ is the entropy density, hence this assumption means that ``kinematic'' bulk viscosity is constant), or $\zeta=$ const., or in case of a conserved charge one could even assume $\zeta\propto n$ (where $n$ is the conserved charge density). Let us note here that entropy density behaves differently depending on the presence of a conserved charge: if there is none, then simply $s\propto T^\kappa$, while if there is one, the $p=nT$ definition can be utilized, leading to $s \propto n$ and a logarithmic temperature dependence. This can be summarized as follows:
\begin{align}
\textnormal{ conserved charge }n &&\rightarrow&& p &= nT &&\rightarrow&\;& s = s_0\frac{n}{n_0} + n\ln\left(\frac{n_0}{n}\frac{T^\kappa}{T_0^\kappa}\right),\\
\textnormal{no conserved charge }&&\rightarrow&& \varepsilon&\equiv\varepsilon(s) &&\rightarrow&\;& s = s_0\left(\frac{T}{T_0}\right)^\kappa.
\label{e:entropy:non}
\end{align}
We will investigate these possibilities in the next section.

In order to focus on bulk viscosity here, let us start from Hubble flow described in Refs.~\cite{Csorgo:2003rt,Csorgo:2003ry}. This is compatible with a wide set of observables~\cite{Csanad:2009wc,Csanad:2014dpa}, even its perturbations were investigated~\cite{Kurgyis:2018sgx}, and Hubble flow is known to form independently of initial conditions in a wide range of possible scenarios~\cite{Chojnacki:2004ec}. Hubble flow is characterized by the flow profile
\begin{align}
u^\mu = \frac{x^\mu}{\tau},
\end{align}
and in this case, all terms with shear viscosity cancel, as well as terms containing thermal conductivity (unless a spatial profile is also assumed). This makes Hubble flow ideal to study the effects of bulk viscosity. Let us here focus on the case, where all thermodynamic quantities depend only on coordinate proper time $\tau=\sqrt{x^\mu x_\mu}$, and in this case a conserved charge density can be given as
\begin{align}
n=n_0\left(\frac{\tau_0}{\tau}\right)^d,
\end{align}
and similarly for the entropy density $s$, if there is no conserved charge.

It turns out that in the above outlined Hubble flow scenario case the $\tau$ dependence of the pressure $p$ is given by the following ordinary differential equation:
\begin{align}
\kappa\frac{dp}{d\tau} + \frac{d(\kappa{+}1)}{\tau}p - \frac{d^2}{\tau^2}\zeta(p,\tau) &= 0.
\label{e:ptaudiff}
\end{align}

\section{New solutions}
As mentioned above, we have to make some assumption on the thermodynamic behavior of bulk viscosity, as well as on the existence of a conserved charge. Taking these into account, let us introduce the following five cases:
\renewcommand{\labelenumi}{(\Alph{enumi})}
\begin{enumerate}
\setlength\itemsep{0.33em}
\item No conserved charge, constant $\zeta$:\\
 \quad $\zeta = \zeta_0\textnormal{ (const)},\quad \varepsilon =\kappa p,\quad  p= p_0(T/T_0)^{\kappa+1}$.
\item With conserved $n$, constant $\zeta$:\\
 \quad $\zeta = \zeta_0\textnormal{ (const)},\quad \varepsilon=\kappa p, \quad p = nT$.
\item No conserved charge, $\zeta{\propto} s$:\\
 \quad $\zeta = \zeta_0(T/T_0)^\kappa, \quad \varepsilon=\kappa p,\quad p= p_0(T/T_0)^{\kappa+1}$.
\item With conserved $n$, $\zeta/n=$const:\\
 \quad $\zeta = \zeta_0(n/n_0), \quad \varepsilon=\kappa p,\quad p = nT$.
\item With conserved $n$, ``$\zeta{\propto} s$'':\nonumber\\
 \quad $\zeta = \zeta_0(T/T_0)^\kappa, \quad \varepsilon=\kappa p,\quad p = nT$.
\end{enumerate}
Not all assumptions lead to physically relevant solutions, as we point out below. Nevertheless, let us summarize the corresponding solutions can be given as follows.\\
Cases A and B (note that the definition of temperature is different in the two cases):
\begin{align}
\quad p(\tau) =& \left[p_0 - \frac{d^2}{(\kappa{+}1)d - \kappa}\frac{\zeta_0}{\tau_0}\right]
\left(\frac{\tau_0}{\tau}\right)^{d\frac{\kappa{+}1}{\kappa}} + \frac{d^2}{(\kappa{+}1)d - \kappa}\frac{\zeta_0}{\tau} .
\end{align}
Case C:
\begin{align}
 p(\tau)= 
\begin{cases}
p_0\bigg\{\left(1+\frac{d^2}{(\kappa{+}1)(\kappa{-}d)}\frac{\zeta_0}{p_0\tau_0}\right)\left(\frac{\tau_0}{\tau}\right)^{\frac d\kappa} - \frac{d^2}{(\kappa{+}1)(\kappa{-}d)}\frac{\zeta_0}{p_0}\frac 1\tau\bigg\}^{\kappa+1}, & \text{for } \kappa\neq d,\\
 p(\tau) = p_0\bigg[1 + \frac{\kappa}{\kappa{+}1}\frac{\zeta_0}{p_0\tau_0}\ln\frac\tau{\tau_0}\bigg]\left(\frac{\tau_0}{\tau}\right)^{\kappa+1}, & \text{for } \kappa=d.
\end{cases}
\end{align}
Case D:
\begin{align}
p(\tau) = 
\begin{cases}
\left[p_0{+}\frac{d^2}{\kappa{-}d}\frac{\zeta_0}{\tau_0}\right]\left(\frac{\tau_0}{\tau}\right)^{\frac{\kappa+1}{\kappa} d}
- \frac{d^2}{\kappa{-}d}\frac{\zeta_0}{\tau_0}\frac{\tau_0^{d+1}}{\tau^{d+1}}, & \text{for } \kappa\neq d,\\
p(\tau) = p_0\left[1 + \frac{\zeta_0\kappa}{p_0\tau_0}\ln\frac{\tau}{\tau_0}\right]\left(\frac{\tau_0}{\tau}\right)^{\kappa+1}, & \text{for } \kappa=d.
\end{cases}
\end{align}
Case E:
\begin{align}
p(\tau) =\!
\begin{cases}
p_0\bigg\{\!\bigg(1{-}\frac{d^2(\kappa{-}1)}{\kappa{-}d}\frac{\zeta_0}{p_0\tau_0}\bigg)\big(\frac{\tau}{\tau_0}\big)^{d\frac{\kappa^2{-}1}{\kappa}}{+}\frac{d^2(\kappa{-}1)}{\kappa{-}d}\frac{\zeta_0}{p_0\tau_0}\big(\frac{\tau}{\tau_0}\big)^{d\kappa{-}1}\!\bigg\}^{-\frac1{\kappa-1}},\!&\! \text{for } \kappa{\neq}d,\\
p_0\big(\frac{\tau_0}{\tau}\big)^{\kappa+1}\bigg\{1 - \kappa(\kappa{-}1)\frac{\zeta_0}{p_0\tau_0}\ln\frac{\tau}{\tau_0}
\bigg\}^{-\frac1{\kappa-1}},\!&\! \text{for } \kappa{=}d.
\end{cases}
\label{e:caseEsol}
\end{align}

\begin{figure}
\begin{center}
\includegraphics[width=1\linewidth]{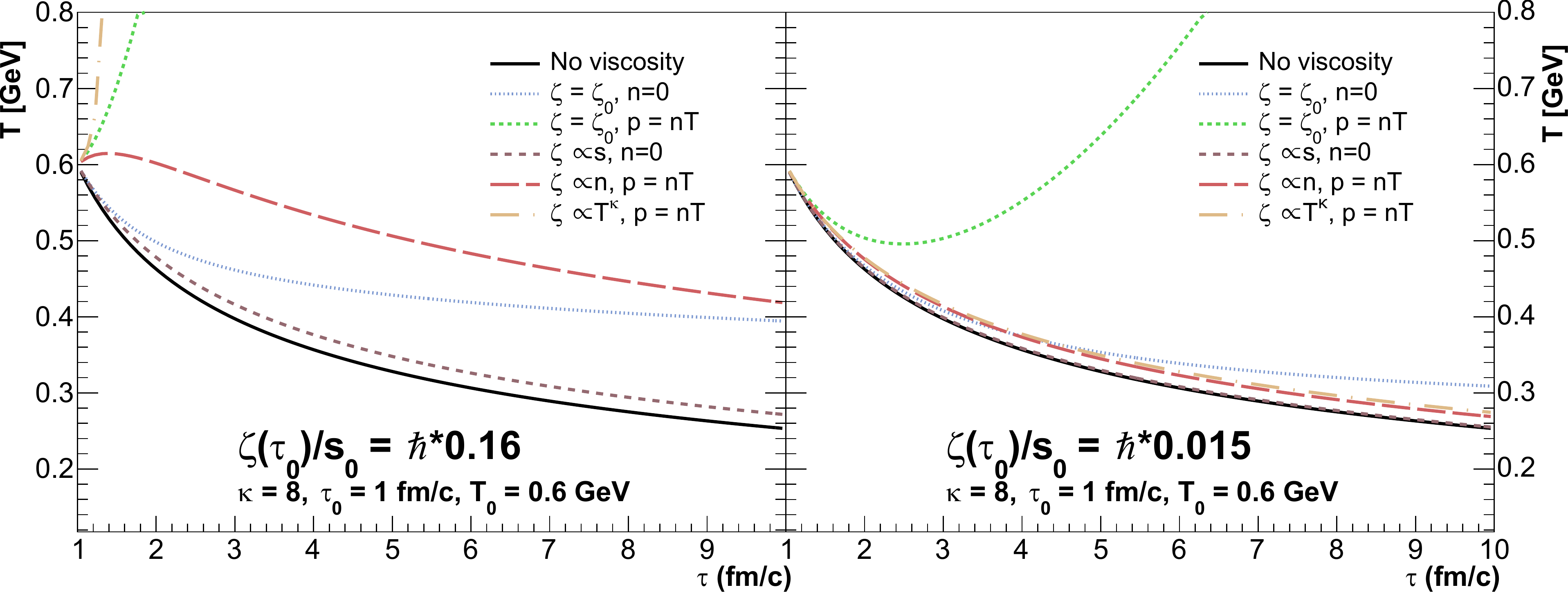}
\end{center}
\caption{Proper time dependence of the temperature in case of the different solutions, for two different viscosity scales. Note that the viscosity changes very differently in each of the cases, hence one figure does not compile different solutions at the same viscosity, just the initial value is the same. The solid line shows the perfect fluid case, which corresponds to vanishing bulk viscosity and the fastest cooling. This curve is the same on both panels. Note furthermore that the third (green short dashed) and last (brown dash-dotted) curves represent the non-physical cases B and E where heating is the dominant process. In the physical cases cooling is dominant for large times.}
\label{f:Ttau_zeta}
\end{figure}

From the above cases, B and E are not physical, since in case B  $T(\tau)$ diverges (because $p=nT$ and $n$ decreases faster than $p$), while in case E even $p(\tau)$ diverges if $d>\kappa$ or if $\frac{d^2(\kappa{-}1)}{\kappa{-}d}\frac{\zeta_0}{p_0\tau_0}>1$ (i.e. even for moderate bulk viscosities). We compare all cases on Fig.~\ref{f:Ttau_zeta} for two different values for $\zeta(\tau_0)/s_0$.  It is clearly visible that cases B and E (corresponding to solutions with $p = n T$ and bulk viscosity constant or proportional to $T^{\kappa}$) are not physically relevant since they result in diverging temperatures, in other words the heat production by the bulk viscosity is so dominant that it leads to a temperature that increases for late times despite the Hubble expansion of the fireball. In the physical cases of A, C and D, the cooling due to the expansion is the dominant process for sufficiently late freeze-out times. It is interesting to note that among the physical cases the largest viscous effect for large times is present in case A. Comparing the two plots with the different bulk viscosity scales, one may furthermore observe that in case D, bulk viscosity may result in an initial reheating if the bulk viscosity is large enough.

\section{Summary and conclusions}
In this paper we presented solutions of viscous hydrodynamics for different assumptions on the evolution of bulk viscosity. These solutions are valid for any value of shear viscosity, since it cancels from the equations in case of the outlined Hubble flow. The different cases correspond to different scenarios, two of them not physical (as they lead to ever increasing temperatures). The others may be utilized in analytically investigating the effect of bulk viscosity on the time evolution of the strongly interacting quark gluon plasma created in high energy heavy ion collisions.

\section*{Acknowledgments}
This research has been partially supported by the NKFIH grants No. FK-123842 and FK-123959, the EFOP 3.6.1-16-2016-00001 grant (Hungary), and THOR, the EU COST Action CA15213. M. Cs. was supported by the J{\'a}nos Bolyai Research Fellowship of the Hungarian Academy of Sciences and {\'U}NKP-19-4 New National Excellence Program of the Hungarian Ministry for Innovation and Technology.

\bibliographystyle{pepan}
\bibliography{../../../Master}

\end{document}